\documentclass{article}
\usepackage[utf8]{inputenc}
\usepackage{authblk}
\usepackage{setspace}
\usepackage[margin=1.25in]{geometry}
\usepackage{graphicx}
\graphicspath{ {./figures/} }
\usepackage{subcaption}
\usepackage{amsmath}
\usepackage[T1]{fontenc}
\usepackage{lineno}
\usepackage{xcolor}
\usepackage{color,soul} 
\usepackage{float}

\usepackage[english]{babel}
\usepackage[sorting=none]{biblatex}
\title{Bose condensation of upper-branch exciton-polaritons in a transferrable microcavity}

\author[1$\dag$]{Xingzhou Chen}
\author[2$\dag$]{Hassan Alnatah}
\author[1]{Danqun Mao}
\author[1]{Mengyao Xu}
\author[2]{Qiaochu Wan}
\author[2]{Jonathan Beaumariage}
\author[1]{Wei Xie}
\author[1]{Hongxing Xu}
\author[1]{Zhe-Yu Shi}
\author[2*]{David Snoke}
\author[1*,3]{Zheng Sun}
\author[1*,3,4,5]{Jian Wu}

\affil[1]{State Key Laboratory of Precision Spectroscopy, East China Normal University, Shanghai, 200241, China.}
\affil[2]{Department of Physics and Astronomy, University of Pittsburgh, Pittsburgh, PA 15260, USA.}
\affil[3]{Collaborative Innovation Center of Extreme Optics, Shanxi University, Taiyuan, Shanxi 030006, China.}
\affil[4]{Chongqing Key Laboratory of Precision Optics, Chongqing Institute of East China Normal University, Chongqing 401121, China.}
\affil[5]{CAS Center for Excellence in Ultra-intense Laser Science, Shanghai 201800, China.}
\affil[*]{Address correspondence to: snoke@pitt.edu; zsun@lps.ecnu.edu.cn; jwu@phy.ecnu.edu.cn}
\affil[$\dag$]{These authors contributed equally to this work.}

\newcommand{\beginsupplement}{
    \setcounter{table}{0}
    \renewcommand{\thetable}{S\arabic{table}}
    \setcounter{figure}{0}
    \renewcommand{\thefigure}{S\arabic{figure}}
    \setcounter{equation}{0}
    \setcounter{section}{0}
    \renewcommand{\theequation}{S\arabic{equation}}
}

\begin{document}

\maketitle

\begin{abstract}
Exciton-polaritons are composite bosonic quasiparticles arising from the strong coupling of excitonic transitions and optical modes. Exciton-polaritons have triggered wide exploration in the past decades not only due to their rich quantum phenomena such as superfluidity, superconductivity and quantized vortices but also due to their potential applications for unconventional coherent light sources and all-optical control elements. 
Here, we report the observation of Bose-Einstein condensation of the upper polariton branch in a transferrable WS$_2$ monolayer microcavity. Near the condensation threshold, we observe a nonlinear increase in upper polariton intensity. This sharp increase in intensity is accompanied by a decrease of the linewidth and an increase of the upper polariton temporal coherence, all of which are hallmarks of Bose-Einstein condensation. By simulating the quantum Boltzmann equation, we show that the upper polariton condensation only occurs for a particular range of particle density. We can attribute the creation of Bose condensation of the upper polariton to the following requirements: 1) the upper polariton is more excitonic than the lower one; 2) there is relatively more pumping in the upper branch; and 3) the conversion time from the upper to the lower polariton branch is long compared to the lifetime of the upper polaritons. 
\end{abstract}

\section{Introduction}
The realization of exciton-polariton strong coupling in semiconductor microcavities from cryogenic temperatures \cite{balili2007bose,kasprzak2006bose,kim2011dynamical} up to room temperature \cite{su2017room,plumhof2014room} not only advances the deeper understanding of the fundamental studies of many-body physics but also opens up opportunities for the exploration of potential all-optical control devices \cite{peter2005exciton,chikkaraddy2016single,feng2021all,amo2010exciton,sun2019}. Exciton-polaritons (called here simply ``polaritons'') are hybridized half-light half-matter quasiparticles arising from the strong coupling of cavity photons and excitonic transitions. In the strong coupling regime, the rate of energy exchange between the cavity photons and the excitons becomes much faster than their dissipation rates, giving rise to new mixed states, known as the upper polariton (UP) and the lower polariton (LP) branches. As interacting bosons, polaritons can exhibit Bose-Einstein condensation, which can be characterized by a strong coupling mechanism, e.g., a nonlinear increase of output strength, narrowing of spectra, continuous blue-shift, an increase of temporal and spatial coherence across the threshold. The dual light-matter nature allows flexible control of the condensation of polaritons and facilitates its potential applications in quantum simulation \cite{luo2020classical,boulier2020microcavity}, unconventional coherent light sources \cite{zhang2015weak,schneider2013electrically}, all-optical polarization logic devices \cite{liew2008optical,liew2010exciton}, neural morphology computing \cite{berloff2017realizing,ballarini2020polaritonic}, and transistors \cite{ballarini2013all,zasedatelev2019room}. 
\par
 Polariton condensates allow the study of a range of behavior as the cavity lifetime is changed. In the limit of long cavity lifetime, the scattering rate from UP to LP is fast, so that the dynamics approach the behavior of a single population model, making the UP hard to detect. As the cavity lifetime is decreased, it becomes comparable to the conversion time from UP to LP, so that the dynamics resemble that of two populations competing for condensation. The experiments we have performed here are in the latter case; the polaritons are in the strong-coupling limit with a lifetime comparable to their conversion time. 
 This opens up new avenues to explore out-of-equilibrium quantum phenomena. 
 A few experiments have explored the upper polariton branch in a variety of microcavities based on different materials, including the transition from the upper polariton state to the dark exciton state \cite{groenhof2019tracking}, terahertz radiation originating from the scattering of upper polaritons to lower polaritons  \cite{kavokin2010stimulated} and upper polaritons participating in parametric  amplification \cite{dasbach2002tailoring}. However, in these studies, the focus has largely been in a regime well below the threshold of condensation.
\par
In the present work, we have successfully demonstrated upper polariton condensation formed in a symmetric transferrable WS$_2$ monolayer microcavity. Through a comprehensive investigation that combines experimental and theoretical approaches, we explored upper polariton condensation resulting from non-resonantly pumped polaritons. Our focus was on understanding the specific conditions necessary for the upper polariton to undergo condensation at a lower threshold than the lower polariton. Our findings have revealed that upper polariton condensation occurs within a particular range of particle density ( < 10\textsuperscript{13}/cm\textsuperscript{2}). This Bose condensation of the upper polariton can be attributed to several crucial factors. This discovery is of great significance as it contributes to the fundamental understanding of out-of-equilibrium quantum phenomena and unlocks new possibilities for further investigations. Additionally, our results hold practical relevance for designing and developing polaritonic lasers, where the competition between condensates of the lower and upper polariton states plays a crucial role.
\par
\begin{figure}
    \centering
    \includegraphics[width=1\textwidth]{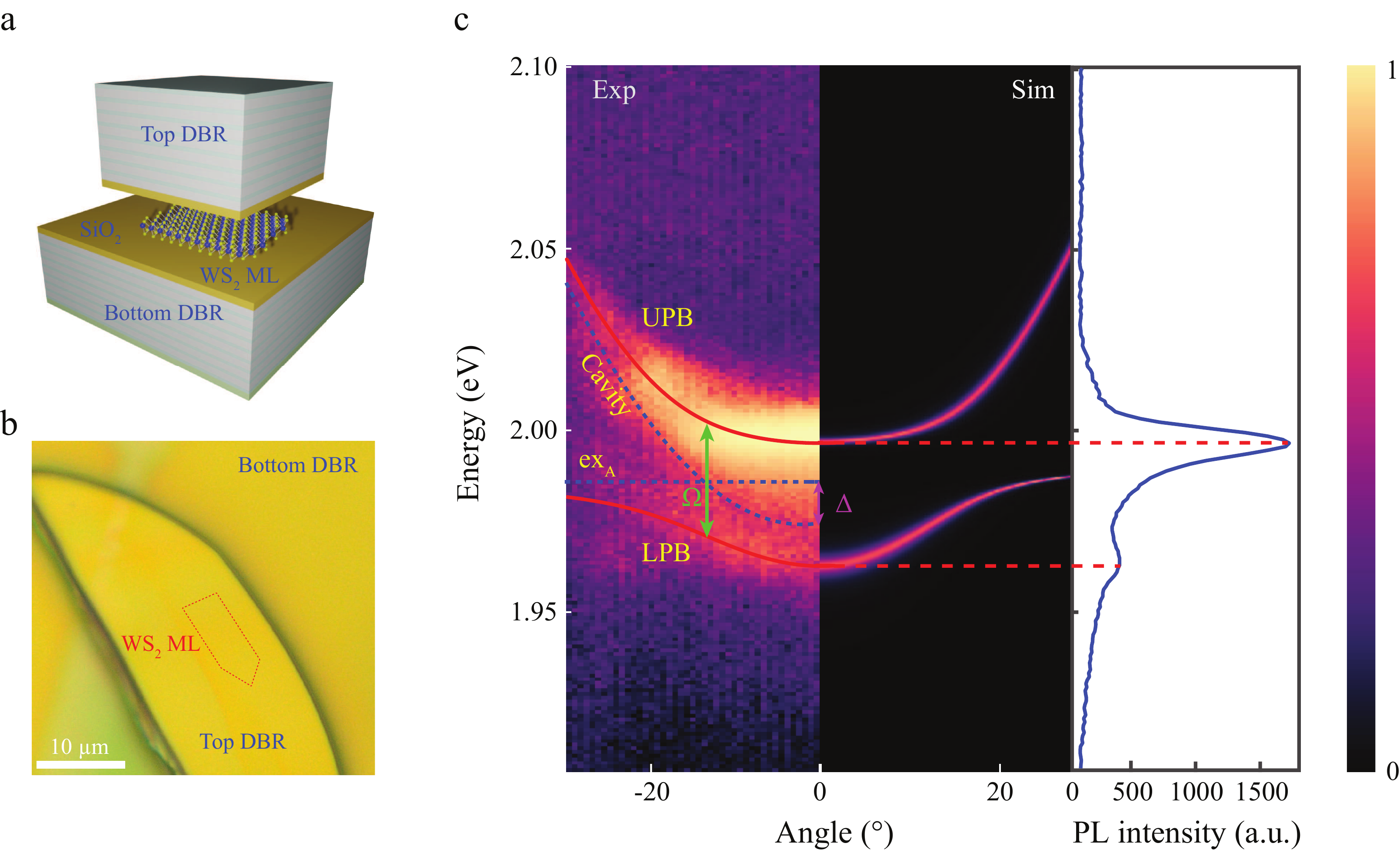}
    \caption{\textbf{Sample structure and optical properties of WS$_2$ monolayer microcavity polaritons.} \textbf{a,} Schematic of the microcavity structure. The bottom DBR is composed of 12 pairs of SiO$_2$/SiN$_x$ grown by PECVD. The top DBR was mechanically separated from the substrate and transferred on top of the WS$_2$ monolayer. \textbf{b,} Optical microscope image of the full microcavity. The dotted red line indicates the WS$_2$ monolayer position. \textbf{c,} Angle-resolved PL measurements (left) compared with the simulated absorption image (middle) of the WS$_2$ monolayer microcavity, showing the same lower (LPB) and upper (UPB) polariton branches. The red lines are fitted to the LPB and UPB dispersion and the blue dotted lines are fitted to the cavity and the ex$_A$ mode with the coupled harmonic oscillator model, giving the Rabi splitting ($\Omega$) of 30 meV and cavity-exciton energy detuning ($\Delta$) of -11 meV, respectively. The strong coupling is illustrated by the PL spectrum (right), showing two dominant peaks assigned as the upper and lower polaritons, denoted with the red dotted lines.}
    \label{fig:1}
\end{figure}

\section{Experimental Design and Results}
The semiconductor-type two-dimensional transition metal dichalcogenides (2D TMDCs) are a group of naturally abundant materials with MX$_2$ stoichiometry, where M is a transition-metal element from group VI (M = Mo, W) and X is a chalcogen (X = S, Se). The most intriguing feature of TMDCs is the emergence of fundamentally distinct electronic and optoelectronic properties as the material transitions from bulk to the two-dimensional limit (monolayer) \cite{mak2010atomically,splendiani2010emerging}. In addition, the 2D TMDCs possess a unique combination of properties, including high oscillator strength \cite{wang2018colloquium}, large binding energy (a few hundreds meV) \cite{chernikov2014exciton,he2014tightly,ye2014probing}, significant nonlinear properties\cite{zhang2021van,shahnazaryan2017exciton}, and compelling structural tunability \cite{susarla2017quaternary,sun2021photoluminescence,jones2013optical}. The strong excitonic effects in TMDCs present opportunities for studying light and matter interactions \cite{sun2021charged,galfsky2016broadband,wu2014control,gan2013controlling,liu2015strong,lundt2016room}. This interaction can be further controlled by embedding the 2D TMDCs into an optical microcavity \cite{sun2017optical,chakraborty2018control}. Figure \ref{fig:1}a illustrates a schematic of the full structure design. The WS$_2$ monolayer is sandwiched between two distributed Bragg reflectors (DBRs).
The bottom DBR is made of alternating layers of SiO$_2$/SiN$_x$ with 12 periods and was grown by Plasma Enhanced Chemical Vapor Deposition (PECVD) with the center of the stopband at 628 nm. A spacer layer of quarter wavelength thickness SiO$_2$ was then grown on top of the bottom DBR to form a $\lambda/2$ cavity. The WS$_2$ monolayer which was mechanically exfoliated from bulk crystal (purchased from HQ) was then dry transferred onto the half cavity. Lastly, the top DBR, similar to van der Waals materials, was peeled off the substrate and transferred on top of the monolayer (see also Methods) \cite{rupprecht2021micro,paik2023high}. The isolated WS$_2$ monolayer with a size of 5×10 $\mu$$m^2$ is circled with the red dashed line, as shown in the optical microscope image in Figure \ref{fig:1}b. To show that the polaritons are in the strong-coupling regime, we made use of angle-resolved imaging at a temperature of 10 K (see Fig. \ref{fig:1}c). The angle has a one-to-one correspondence with in-plane momentum (k$_\parallel$) through the relation k$_\parallel$= ($\omega$/c)sin$\theta$, where c is the speed of light. A pump laser at 515 nm with a repetition frequency of 76 MHz was used to efficiently excite the sample (see also Methods). Figure \ref{fig:1}c unambiguously shows the anti-crossing feature of the dispersion illustrated together with the simulated absorption intensity (middle) showing the LP and UP branches via the transfer matrix method. The red lines are fits to the LP and UP dispersions and the blue dotted lines are fits to the cavity and the ex$_A$ mode using a coupled harmonic oscillator model, giving a Rabi splitting ($\Omega$) of 30 meV and cavity-exciton energy detuning ($\Delta$) of -11 meV, respectively. The PL spectrum further shows the strong coupling by summing for a narrow range of emission angles around k$_\parallel$ = 0. The two dominant peaks in Figure \ref{fig:1}c are assigned as the upper and lower polariton peaks, denoted with the red dotted lines. Crucially, the upper polariton branch shows a brighter emission than that of the lower polariton branch in this sample.
\par
\begin{figure}[!ht]
    \centering
    \includegraphics[width=1\textwidth]{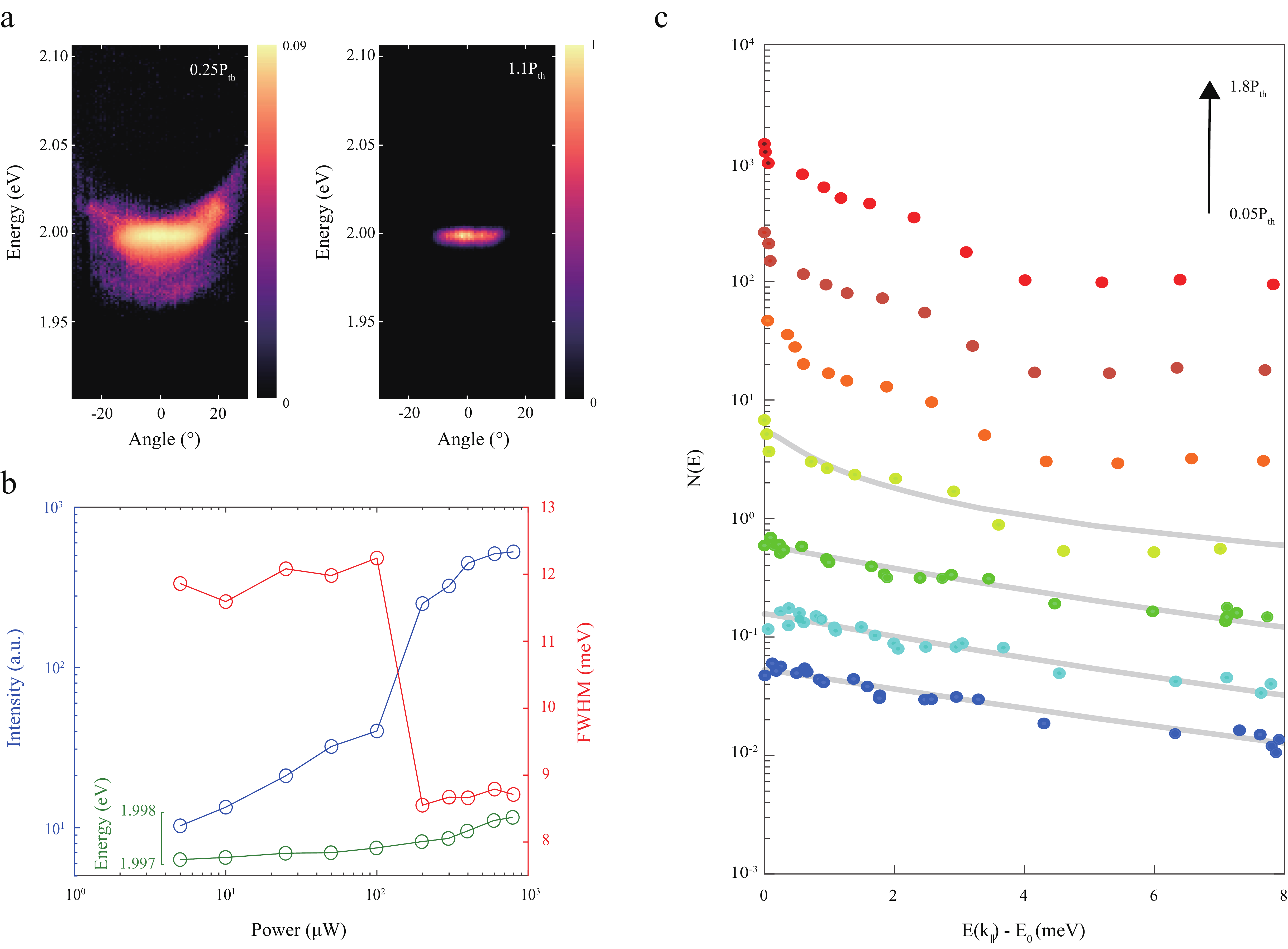}
    \caption{\textbf{Power-dependent measurements at 10 K.} \textbf{a,} Angle-resolved PL measured at 0.25P$\textsubscript{th}$ (left panel) and 1.1P$\textsubscript{th}$ (right panel), where P$\textsubscript{th}$ = 110 $\mu$W is the threshold power of the condensation. Below the threshold, the emission is broadly distributed in momentum and energy in UPB and LPB. Above the threshold, the emission comes almost exclusively from the k$_\parallel$ = 0 lowest energy state from the UPB. \textbf{b,} Occupancy of the k$_\parallel$ = 0 ground state (blue circles), linewidth (red circles) and energy blueshift (green circles) versus the pump power for the upper polariton branch. At low pump power, the ground state occupancy increases linearly with the excitation and then increases exponentially after the threshold before becoming linear again. \textbf{c,} Upper polariton occupation expressed in a semi-logarithmic scale for various pump powers. The power values from low to high are 0.05, 0.2, 0.9, 1.1, 1.3, 1.5, and 1.8 times the threshold values. The solid curves are best fits obtained from the solution of the Boltzmann equation. The fitted temperature (T) is 89 K and the reduced chemical potential $(\mid \mu/k\textsubscript{B}T\mid)$ is 0.168 for the excitation power 1.1P$\textsubscript{th}$, indicating the polariton gas is in the degenerate regime. For exciton power larger than 1.1P$\textsubscript{th}$, the experimental occupation becomes non-thermal, while our model predicts thermal equilibrium. Possible reasons for this are discussed in the main text.}
    \label{fig:2}
\end{figure}

The upper polariton condensation is observed as we increase the pump power across the phase transition, as shown in Figure \ref{fig:2}a-b. At low pump power, the upper polariton occupation increases linearly with power. Near the condensation threshold, a nonlinear increase in intensity is observed, which becomes linear again at much higher pump power. This sharp increase in intensity is accompanied by a decrease of the linewidth by a factor of 4, energy blueshift of $\sim$ 1 meV, that is 3.3\% of the Rabi splitting, and an increase of the polariton temporal coherence, all of which are hallmarks of Bose-Einstein condensation. The saturation and the slight broadening of the linewidth above the threshold can be attributed to the increased polariton-polariton interaction. To get more insight into the quantum statistics of the upper polariton, we make use of angle-resolved imaging, giving the intensity  $I\left ( k,E \right )$, which is then converted to an occupation number $N\left ( E \right )$ using a single efficiency factor (for more details, we refer the reader to Reference \cite{sun2017bose}). The measured UP occupation for different pump power values is shown in Figure \ref{fig:2}c, where we have defined the ground-state energy in each case as $E = 0$.
\par
We find that when the pump power is below the condensation threshold, the occupation of the upper polaritons is well described by a Maxwell-Boltzmann distribution, indicating that the polariton gas is in thermodynamic equilibrium with a well-defined temperature. When the pump power is above the threshold, we observe a significant increase in the upper polariton ground state occupation. However, the high-energy tail becomes non-thermal. With increasing pump power, this feature maintains the same behavior in the distribution, consistent with the other reported experiments carried out in short lifetime polariton systems demonstrating nonequilibrium polariton condensates \cite{kasprzak2006bose,kasprzak2008formation}. Slightly above the threshold ($\sim$1.1P$\textsubscript{th}$), the upper polariton occupation is quasi-thermal, allowing us to get reasonable fits to a Bose-Einstein distribution as shown in Figure \ref{fig:2}c. The fitted value of T is 89 K and the corresponding chemical potential is -1.3 meV for the fourth occupation curve in Fig.~\ref{fig:2}, leading to a reduced chemical potential $\mid \mu/k\textsubscript{B}T\mid$ = 0.168, which indicates the quantum degeneracy of upper polariton gas under such pump power.
\par
\begin{figure}[!ht]
    \centering
    \includegraphics[width=1\textwidth]{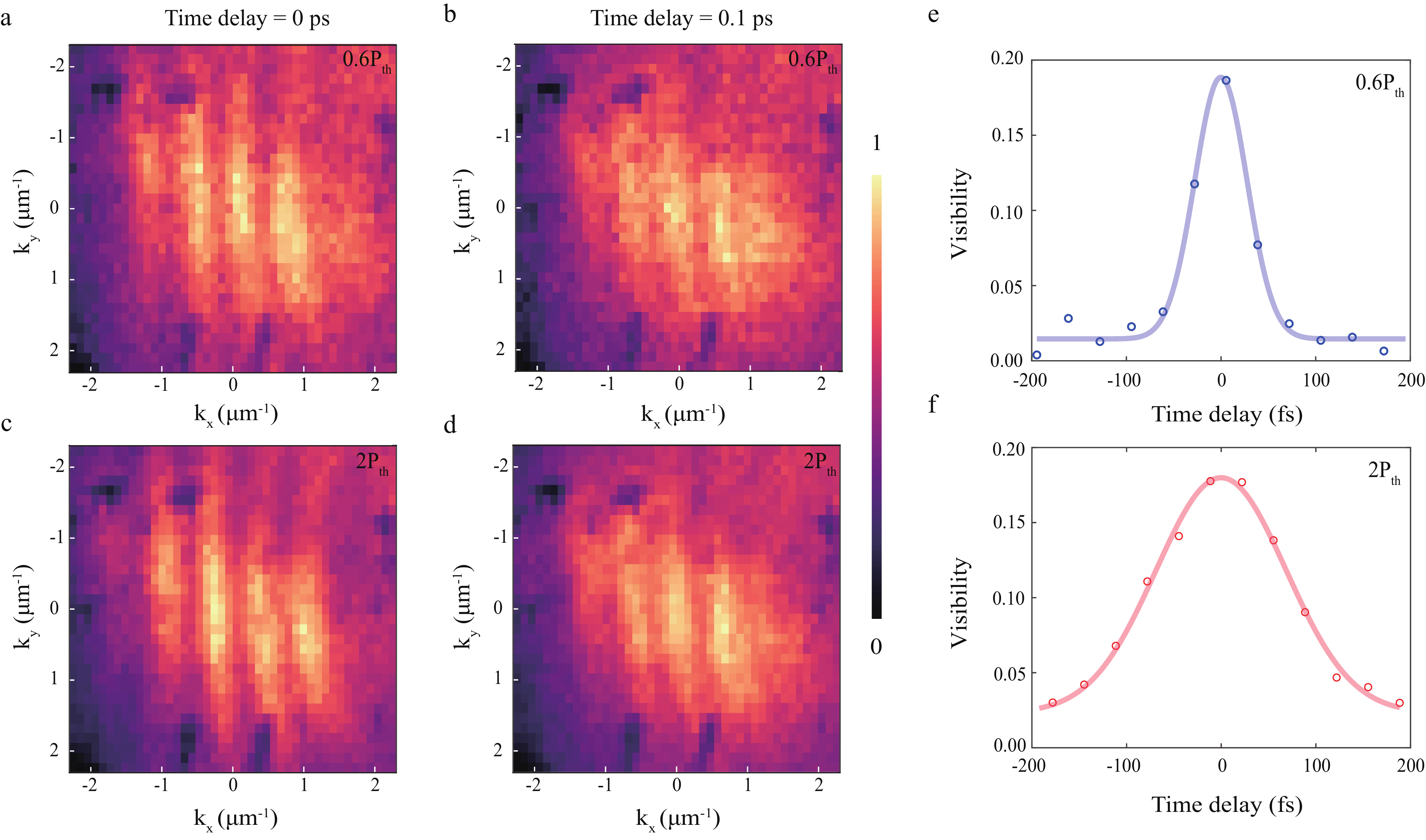}
    \caption{\textbf{Temporal coherence for upper polaritons.} \textbf{a, b,} Typical interference patterns pumped below and \textbf{c, d,} above the threshold with two different time delays, $\Delta$t = 0 ps (left column) and $\Delta$t = 0.1 ps (right column), respectively. \textbf{e, f,} Visibility as a function of the time delay for the corresponding pump powers. The experimental data were fitted with a Gaussian function. The coherence time is increased from $\delta$t = 55 fs to $\delta$t = 138 fs when the excitation power increases from 0.6P\textsubscript{th} up to 2P\textsubscript{th}.}
    \label{fig:3}
\end{figure}

To confirm the extended coherence expected of the condensation for upper branch exciton-polariton, we studied the temporal coherence using a Michelson interferometer. By tuning the relative distance between the two arms of the Michelson interferometer, we can introduce a time delay to study the temporal coherence of the upper polariton gas. We applied a conjugated lens to convert real space to momentum space.  Typical interference patterns below (0.6P\textsubscript{th}) and above the threshold (2P\textsubscript{th}) are shown in Figure \ref{fig:3}a-d, with two different time delays, $\Delta$t = 0 ps (left column) and $\Delta$t = 0.1 ps (right column) respectively. The interference patterns without time delay exhibit well-contrasted interference fringes indicating the emergence of extended coherence in the system. As we introduced a time delay, we see a reduction in the visibility allowing us to extract the coherence time. Here the visibility is defined as the ratio of the intensity difference of the maximum and minimum and the sum of them (I\textsubscript{max}-I\textsubscript{min})/(I\textsubscript{max}+I\textsubscript{min}). In Figure \ref{fig:3}e-f, we plot the visibility as a function of time delay. To determine the coherence time, the experimental visibility data were fitted with a Gaussian function. We find that the coherence time increased from $\delta$t = 55 fs to $\delta$t = 138 fs when the excitation power increased from 0.6P\textsubscript{th} to 2P\textsubscript{th}.
\par

In addition, to further understand the underlying mechanism of the condensation in the upper polaritons, we carried out power-dependent dynamic measurements using a streak camera with a typical temporal resolution of 1 ps, as shown in Figure \ref{fig:4}a, c.
The time-resolved photoluminescence spectra for UP/LP were extracted using a two-peak fitting procedure. These time-resolved spectra allow us to extract several crucial parameters by fitting them to a population dynamics model as shown in Figure \ref{fig:4}b, d
(see next section for details). 
\par

\section{Theoretical Modelling and Discussion}
\par
To model the dynamics of the polaritons, we used two different models---a coarse-grained model to describe the evolution of the total polariton density as a function of time $n(t)$, and a fine-grained model, namely, a full quantum Boltzmann equation solution, to describe the evolution of the polariton density as a function of both time and energy, $n(t,E)$. The coarse-grained model allows us to extract parameters such as the conversion time between the UP and LP and characteristic lifetime parameters of the system, while the fine-grained model allows us to extract the occupation of the polaritons as a function of their energy. 
\par
First, we discuss the coarse-grained model. We use the following dynamic equations to simulate the conversion between the reservoir, the upper and the lower polaritons:
\begin{align}
\frac{dn_\text{res}}{dt}&=-\frac{n_\text{res}}{\tau_{\text{res}}}-K_\text{u}n_\text{res}^2-K_\text{l}n_\text{res}^2\nonumber\\
\frac{dn_\text{up}}{dt}&=+\frac{n_\text{res}}{\tau_\text{up}}-\frac{n_\text{up}}{\tau_\text{up}}-\frac{n_\text{up}}{\tau_\text{conv}}+\frac{n_\text{lp}e^{-\Delta/k_\text{B}T}}{\tau_\text{conv}}+K_\text{u}n_\text{res}^2\nonumber\\
\frac{dn_\text{lp}}{dt}&=+\frac{n_\text{res}}{\tau_\text{lp}}-\frac{n_\text{lp}}{\tau_\text{lp}}+\frac{n_\text{up}}{\tau_\text{conv}}-\frac{n_\text{lp}e^{-\Delta/k_\text{B}T}}{\tau_\text{conv}}+K_\text{l}n_\text{res}^2.\label{density_dynamics}
\end{align}
Here $n_\text{res}$, $n_\text{up}$ and $n_\text{lp}$ stand for the densities of reservoir excitons (i.e., excitons with momentum or spin states that do not couple to light) and polaritons in both upper and lower branches, respectively; $\tau_\text{res}$, $\tau_\text{up}$ and $\tau_\text{lp}$ are the corresponding lifetimes. We also introduce two-particle collisional scattering from the reservoir to both polariton branches represented by $K_\text{u}$ and $K_\text{l}$. The conversion between the upper and lower polaritons involves both single-particle decay (e.g., phonon emission) and two-particle scattering. The conversion time $\tau_\text{conv}$ thus depends on the densities of both branches as well as the reservoir, and is written as $\tau_\text{conv}^{-1}=\tau_0^{-1}+A(n_\text{up}+n_\text{lp})+Bn_\text{res}$, where the first term on the RHS describes single-particle decay process, and the rest describe the two-particle scatterings. Note that we have also introduced a Boltzmann factor $e^{-\Delta/k_\text{B}T}$ in the conversion process in the dynamic equation, such that the densities of upper and lower polaritons reach the Boltzmann distribution if other loss processes are ignored.
\par
\begin{figure}[!ht]
    \centering
    \includegraphics[width=0.8 \textwidth]{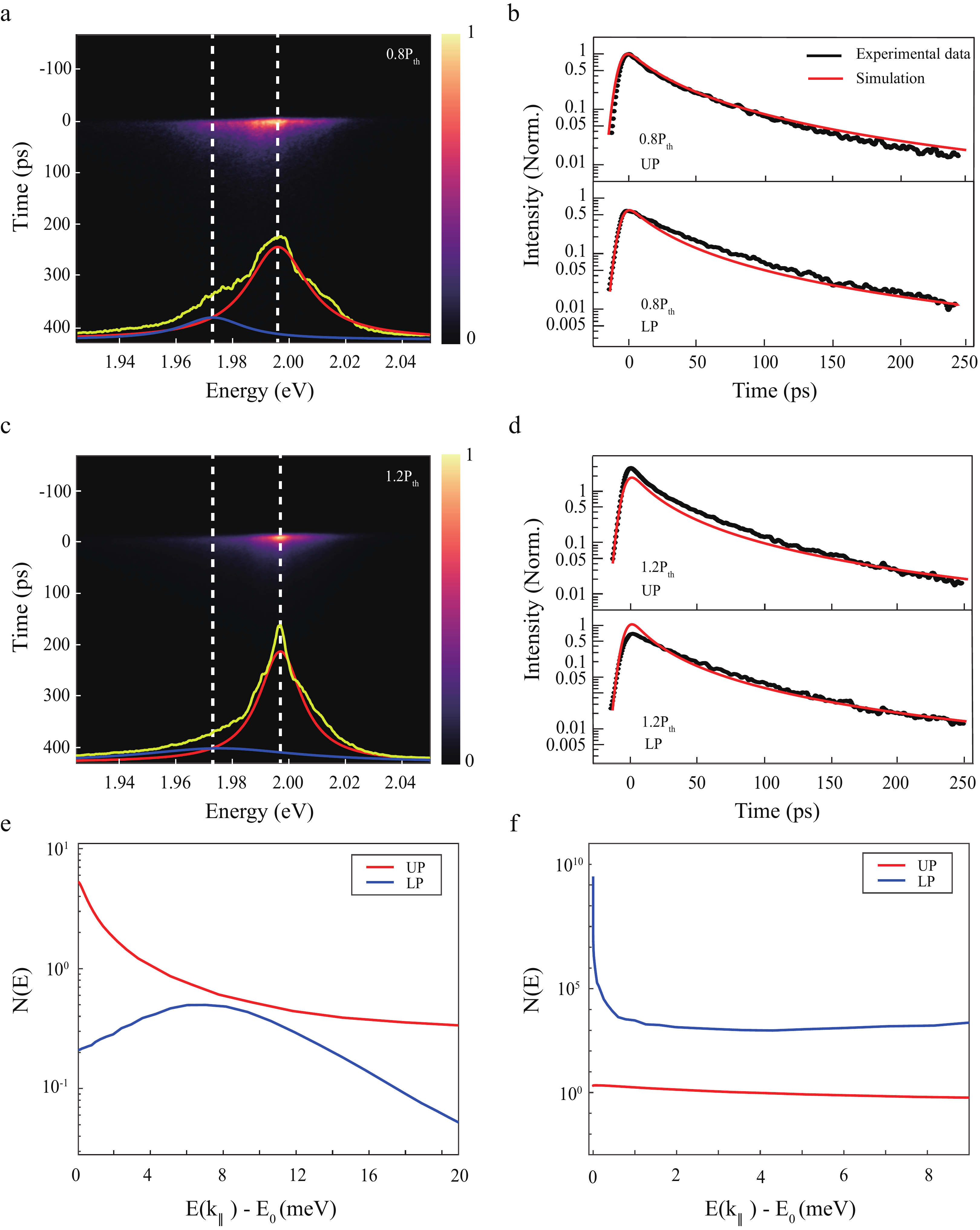}
    \caption{\textbf{Time-resolved photoluminescence spectra and the underlying mechanism for the upper polariton condensate.} \textbf{a, c,} Time-resolved photoluminescence images for the pumping power at 0.8P$\textsubscript{th}$ and 1.2P$\textsubscript{th}$ measured by a streak camera, respectively. The time-integrated spectra (yellow) were fitted by two Lorenz peaks and were assigned to the upper (red) and lower (blue) polaritons. \textbf{b, d,} Experimental measurements (black) and simulated results (red), based on dynamic equations mentioned in the main text and methods) of the time-resolved upper and lower polariton photoluminescence spectra for different pumping power (\textbf{b}: $0.8P_\text{th}$, \textbf{d}: $1.2P_\text{th}$). \textbf{e,} The occupation of the lower and upper polaritons obtained from the quantum Boltzmann simulations for the case of $\frac{P^\mathrm{UP}}{P^\mathrm{LP}} = 2.5$. For this pumping ratio, the upper polariton undergoes condensation at a lower threshold power than the lower one. \textbf{f,} The transition from UP to LP condensation at higher power with the same ratio of pumping $\frac{P^\mathrm{UP}}{P^\mathrm{LP}} = 2.5$.} 
    \label{fig:4}
\end{figure}

 A comparison between the simulations and the time-resolved photoluminescence spectra is shown in Figure \ref{fig:4}b,d. 
 Although there are multiple fit parameters, the fits are highly constrained because we use the same equations to fit multiple initial pumping conditions for both polariton branches over the full range of temporal evolution.
To compare the predictions of dynamic equations~\eqref{density_dynamics} with the time-resolved spectra, we assume that at $t=0$, there is no occupation in both polariton branches, while $n_\text{res}(0)=cP$, with $c$ being the pumping power-reservoir transfer rate. The time-resolved populations $n_\text{up}$ and $n_\text{lp}$ are then found by iterating the differential equations (\ref{density_dynamics}). By fitting the simulated spectra with the experimental data as shown in Figure \ref{fig:4}b,d, we extract the following dynamic parameters: $\tau_\text{up}=1.9\text{ ps}$, $\tau_\text{lp}=0.9\text{ ps}$, $\tau_0=30\text{ ps}$, $\tau_\text{res}=100\text{ ns}$, $c=3.4\text{ $\mu$W}^{-1}$, $K_\text{u}=4.4\times10^{-3}\text{ $\mu$m${}^2\cdot$ps}^{-1}$, $K_\text{l}=8.8\times10^{-3}\text{ $\mu$m${}^2\cdot$ps}^{-1}$, $A=2.9\times10^{-3}\text{ $\mu$m${}^2\cdot$ps}^{-1}$, $B=1.5\times10^{-3}\text{ $\mu$m${}^2\cdot$ps}^{-1}$. We find that the lifetimes of upper/lower polaritons extracted from the fits are very close to the expected lifetimes for upper and lower polaritons at $\vec{k}=0$ based on the $Q$ of the cavity and the Hopfield coefficients for the photon fraction of each. Importantly, we find that the conversion time $\tau_\text{conv}$ that describes the conversion process between the upper and lower polaritons is around $20-30\text{ ps}$ at low density, which is much longer than both $\tau_\text{up}$ and $\tau_\text{lp}$. This indicates that the conversion between the upper and lower polariton branches due to spontaneous and two-particle collisions may be ignored within the time interval of interest, and, more importantly, that it is possible for the upper polaritons to reach thermal (quasi) equilibrium without decaying to the lower branch. However, this conversion time becomes faster as the polariton density increases due to stimulated scattering---eventually becoming fast enough that it is comparable to the lifetime of the polaritons. 
\par
Next, we discuss the fine-grained model, which uses the semiclassical Boltzmann equation to describe the dynamics of both lower and upper polariton branches. In general, the Boltzmann equation has the form,
\begin{equation}
\begin{split} 
\frac{\partial n^{\mathrm{LP}}_{\vec{k}}}{\partial t}&=P^{\mathrm{LP}}_{\vec{k}}(t)-\frac{n^{\mathrm{LP}}_{\vec{k}}}{\tau^{\mathrm{LP}}_{\vec{k}}}+\sum_{\vec{k}^{\prime}} W_{\vec{k}^{\prime} \rightarrow \vec{k}}^{(i)}(t)-\sum_{\vec{k}^{\prime}} W_{\vec{k} \rightarrow \vec{k}^{\prime}}^{(i)}(t)\\
\frac{\partial n^{\mathrm{UP}}_{\vec{k}}}{\partial t}&=P^{\mathrm{UP}}_{\vec{k}}(t)-\frac{n^{\mathrm{UP}}_{\vec{k}}}{\tau^{\mathrm{UP}}_{\vec{k}}}+\sum_{\vec{k}^{\prime}} W_{\vec{k}^{\prime} \rightarrow \vec{k}}^{(i)}(t)-\sum_{\vec{k}^{\prime}} W_{\vec{k} \rightarrow \vec{k}^{\prime}}^{(i)}(t)
\end{split}
\end{equation}
where ${n_{\vec{k}}}$ is the occupation number, $\tau_{\vec{k}}$ is the characteristic lifetime, $P_{\vec{k}}$ is the pumping term and the label $\mathrm{LP}$ and $\mathrm{UP}$ refer to the lower and upper polariton branches respectively. The $W^{(i)}$ are the interaction terms to be considered within each population. Our model includes four types of interactions: polariton-polariton interactions within the same branch, polariton-polariton interactions between different branches (i.e, upper polariton-lower polariton interaction), polariton-phonon interactions, and polariton-electron interactions (there are always typically some free carriers in these structures due to charged impurities). The details of the interactions considered in the model are discussed in the supplementary material. This model is very similar to that used in Ref. \cite{hartwell2010numerical}, but crucially, our model treats the upper and lower polariton populations separately and accounts for conversions between these two populations via optical phonons. 
\par
Figure \ref{fig:2}c shows the fits to the upper polariton occupation at low density measured using the quantum Boltzmann equation. In agreement with the experimentally measured occupation, our model shows that at densities well below the condensation threshold, the upper polariton occupation becomes well described by a Maxwell–Boltzmann distribution. Since the interaction processes scale as the square of the density, the thermal equilibrium seen experimentally at low density means that the polariton interactions are dominated by scattering with free charged carriers, as was also found in Ref. \cite{hartwell2010numerical}. We find that thermalization at low density can only be reproduced by including a background free electron gas with a density of around $10^{10}\;\text{cm}^{-2}$. 

\par When quantum statistics become important (i.e $N\left ( E_{0} \right )\sim 1$), the shape of the distribution changes and an upturn at $E = 0 $ appears, indicating the build-up of the ground state occupation. At very high density, the experimental UP occupation becomes highly non-thermal, while our model predicts that the UP should remain in thermal equilibrium, hence the absence of the fit to the high-density data in Fig~\ref{fig:2}c. We speculate that the deviation from the thermal equilibrium at high density in the experiment is due to two reasons. First, the optical system used in these experiments was polarizing, but the polarization of the condensate was not tracked, which can give rise to misleading $N(E)$ values for the coherent part of the gas. Second, at high density rapid exciton-exciton annihilation can take place due to Auger recombination, which can lead to nonequilibrium. Our quantum Boltzmann model does not account for such processes.  
\par
The UP undergoing Bose-Einstein condensation before the LP can also be shown by our model. When the upper polariton is more excitonic than the lower polariton, it becomes more strongly interacting than the lower polariton, therefore enhancing the relaxation dynamics towards condensation. This is because the polariton-polariton interaction strength is proportional to the square of the exciton fraction. Since the upper polariton is 0.7 excitonic and the lower polariton is 0.3 excitonic in our sample, the upper polariton is approximately an order of magnitude more interacting. In addition, because the upper polariton is more excitonic, it has a longer lifetime than the lower polariton as the lifetime is inversely proportional to the photon fraction. Another important parameter is the ratio of the pumping term ${P^\mathrm{UP}}/{P^\mathrm{LP}}$. In the model, we have assumed that the pumping term is independent of $k$ in each branch. We find that the upper polariton condensation can only happen at a lower threshold pumping power than the lower polariton when ${P^\mathrm{UP}}/{P^\mathrm{LP}} > 2.5$ as shown in Figure \ref{fig:4}e. The ratio of these pumping terms cannot be extracted experimentally, but our model can give us a lower bound of this ratio. Since the non-resonant pump in the experiment is at a higher energy than both the lower and upper polaritons, we expect that more upper polariton states are initially occupied than lower polariton states, which is in agreement with our simulations. Lastly, we mention that upper polariton condensation can only happen if the conversion time between the upper polariton to lower polariton is longer than or comparable to the lifetime of the polaritons. In the case of long-lifetime polaritons, the upper polariton population can decay much faster to lower polariton states than decay outside of the cavity. Because of this, almost all the upper polaritons first convert to lower polaritons before escaping the cavity, making the upper polariton branch hard to detect, as seen in long-lifetime GaAs microcavity experiments\cite{sun2017bose}.
\par
At very high pumping power, our model predicts a second transition from UP condensation to LP condensation. As the density of the polaritons increases, stimulated scattering from the UPB to the LPB becomes significantly enhanced. This in turn leads to a sudden "collapse" of the upper polariton particles to the lower polariton branch leading to LP condensation with a very large occupation at $k = 0$ (See Fig. \ref{fig:3}f). The threshold for such a transition is predicted to occur at a much high power than we can observe experimentally. Another challenge is that at such high power, runaway heating might kick in first, making seeing this transition difficult. 
\par

\section{Conclusions}
In this work, we have shown nonequilibrium Bose condensation of polaritons in the UP branch in a transferable WS$_2$ monolayer microcavity. We showed that the upper polariton gas is quasi-thermal at high density, while it has a well-defined temperature at low density. Using  Michelson interferometry, we showed that the polariton gas had extended coherence in k-space with a coherence of 138 fs at two times the threshold. We modeled the system using the semi-classical Boltzmann equation and population dynamics equations which include both lower and upper polariton branches. Our models shed insight on the conditions needed to achieve condensation of the upper polariton. The small lifetime of the upper polariton compared to the conversion time together with the excitonic fraction of the upper polariton plays an important role in observing this effect. In essence, our work paves the way for further understanding condensation competition in open-dissipative quantum systems while linking to practical applications, shedding light on both fundamental aspects of quantum physics and practical laser technologies.

\section{Acknowledgment}
\begin{sloppypar}
The work was supported by the National Key Research and Development Program of China (2021YFA1200803); National Natural Science Foundation of China (12174111, 12227807); Shanghai Pujiang Program (21PJ1403000); Shanghai Sailing Program (20YF1411600);  D. S. and H. A., J. B. and Q. W.  acknowledge support from U.S. National Science Foundation (DMR-2004570) and MURI (W911NF-17-1-0312).

\section{Author contributions}	
Z. S., D. S. and J. W. supervised the project. XZ. C., DQ. M. and Z. S., fabricated the full samples and performed the optical measurements. H. A., MY. X., ZY. S. and D. S. performed the theoretical modeling. All the authors contributed to the discussion of the results and writing of the manuscript.

\section{Competing financial interests}	
The authors declare no competing financial interests.

\section{Methods}
\textbf{Sample Preparing}
\par
The microcavity sample is in a symmetric structure consisting of 12 periods of SiO$_2$/Si$_3$N$_4$ bottom DBR and 12 periods of top DBR fabricated by Plasma Enhanced Chemical Vapor Deposition (PECVD). Later, a quarter wavelength thickness spacers were pre-grown on both DBRs. Then a WS$_2$ monolayer was firstly transferred onto the bottom DBR by a dry-transfer method. The top DBR flake was peeled from the substrate and transferred onto the WS$_2$ monolayer by a pick-up transfer method. The details of the transfer process can refer to the supplementary information.
\par
\noindent
\textbf{Optical Characterization}
\par
The angle-resolved photoluminescence was measured by imaging the Fourier plane excited by a pulse laser of 515 nm with a repetition frequency of 76 MHz (Light Conversion Pharos). The sample was loaded in a closed-cycle high-vacuum dewar (Montana Instruments cryostation-C2) and cooled down to 10K. A 50$\times$ microscope objective (NA = 0.5) was employed in the measurement. The Fourier image was guided to the spectrometer (Horiba iHR550) and onto the charge-coupled device (CCD) (Horiba 1024×256-BD). The time-resolved PL measurements were obtained by a streak camera with a picosecond-order resolution (Hamamatsu C10910). The temporal coherence measurements were collected using a Michelson interferometer with two mirrors in each arm and one equipped with a one-dimension translation stage to change the delay time.
\end{sloppypar}
\clearpage
\title{\LARGE Supplementary information for: Bose condensation of upper-branch exciton-polaritons in a transferrable microcavity}
\maketitle
\beginsupplement
\subsection*{S1. Standard of procedure of making transferrable microcavity}

\begin{figure}[!ht]
    \centering
    \includegraphics[width=0.8\textwidth]{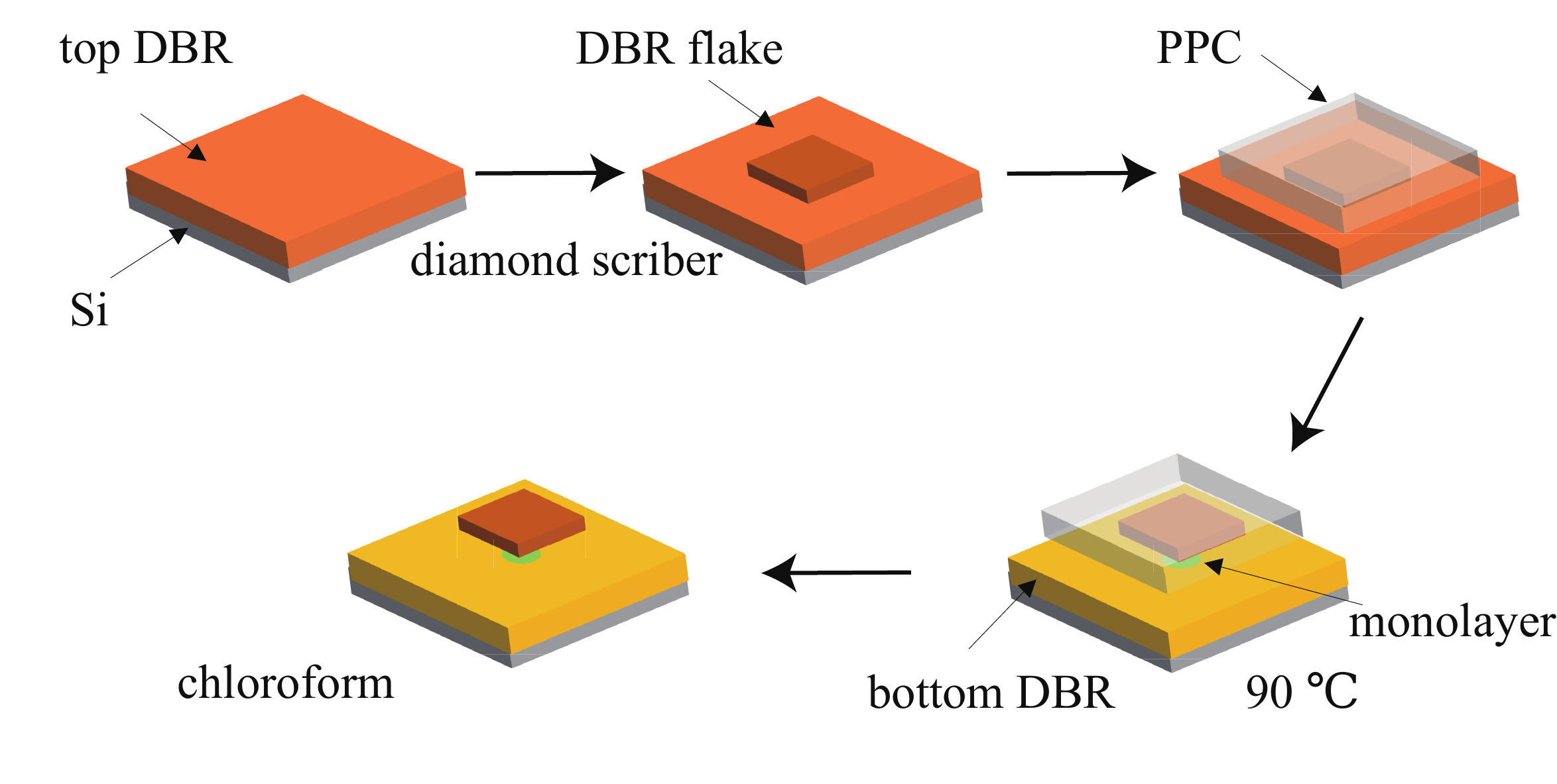}
    \caption{Schematic illustration of the top DBR transfer process. The DBRs were prepared and deposited onto the substrates. Then a monolayer was transferred onto the bottom DBR. A sharp tool (like a diamond scriber) was used to scratch the surface of DBR at moderate pressure, and thereafter the segment of the top DBR was separated from the substrate. We used the polypropylene-carbonate (PPC) film to pick up the DBR flakes at room temperature. After that, we stacked the DBR flake onto the monolayer and released the DBR flake and PPC at 90$^\circ C$. Finally, the PPC was removed in chloroform to leave the DBR flake in the sample.}
    \label{fig:S1}
\end{figure}

\subsection*{S2. Angle-resolved reflectivity spectra of the transferrable empty cavity}
\begin{figure}[!ht]
    \centering
    \includegraphics[width=0.7\textwidth]{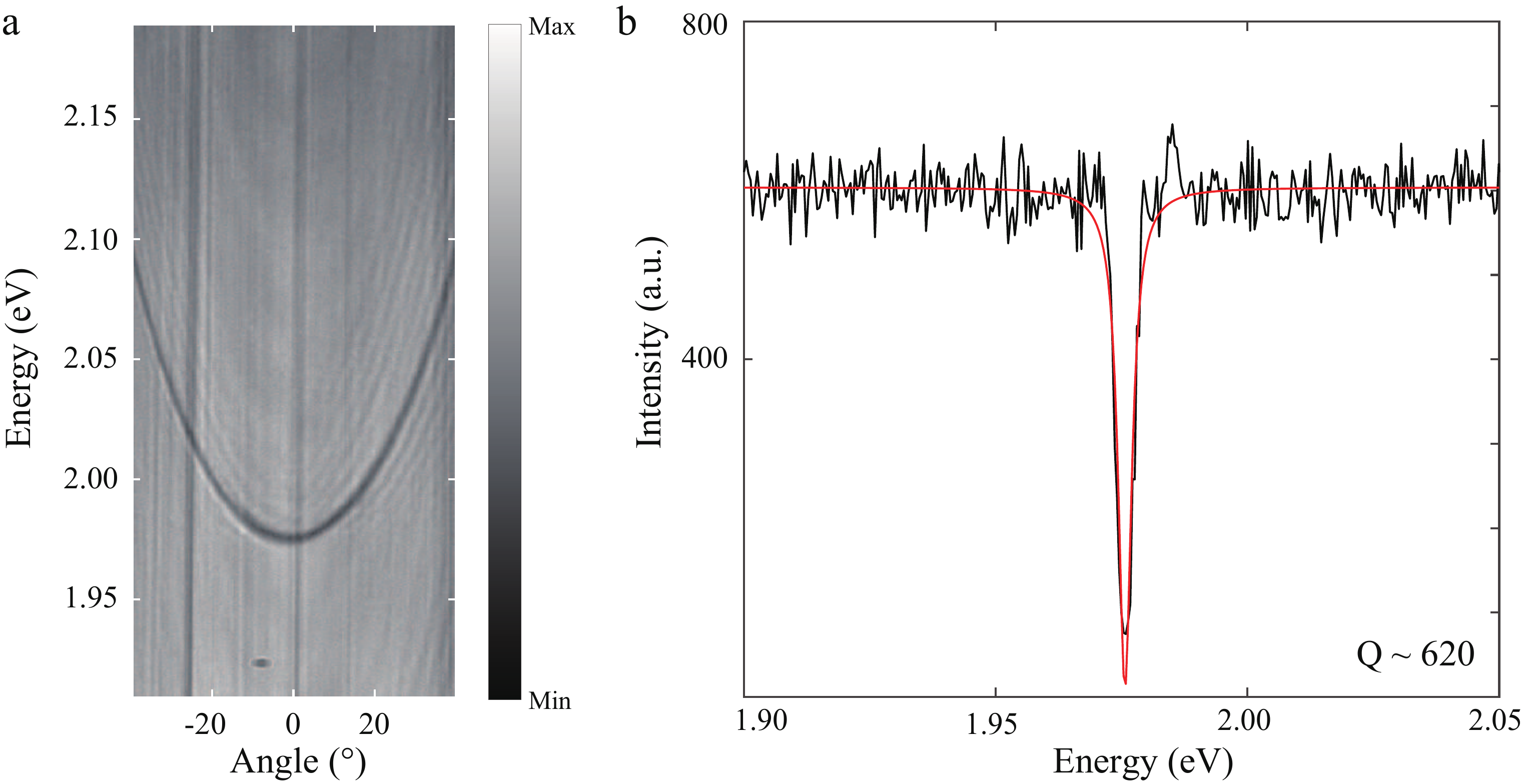}
    \caption{(a) Angle-resolved reflectivity image for the empty microcavity. (b) The reflectivity spectrum was plotted by stripping the pixels along the energy axis near the angle of 0\textdegree, and the Q factor was determined about 620 by fitting the line with the Lorenzian model.}
    \label{fig:S2}
\end{figure}

The Q factor of the transferrable microcavity can be determined by taking the angle-resolved reflectivity measurements. We binned the pixels near the angle = 0\textdegree and plotted the corresponding reflectivity spectrum as the function of the energy presented in Figure S \ref{fig:S2}. The Lorenzian model was exploited to fit the spectrum and gave the Q factor 620. The lifetime of the cavity photon can be calculated at about 0.42 ps by the equation of $\tau$ = $\frac{2Q\hbar}{E}$, where the E is the energy of the cavity photon at angle = 0\textdegree. Lastly, the lifetime of upper polaritons and lower polaritons can be estimated considering the Hopfield coefficient in this sample, that is $\tau_\textsubscript{up}$ = $\frac{\tau_\textsubscript{ph}}{\mid X\mid^2}$ = 1.4 ps and $\tau_\textsubscript{lp}$ = $\frac{\tau_\textsubscript{ph}}{\mid C\mid^2}$ = 0.6 ps, Where the $\mid X\mid^2$ and $\mid C\mid^2$ are the corresponding exciton and cavity photon fraction for the upper polaritons.

\subsection*{S3. Temporal coherence setup}
\begin{figure}[!ht]
    \centering
    \includegraphics[width=0.7\textwidth]{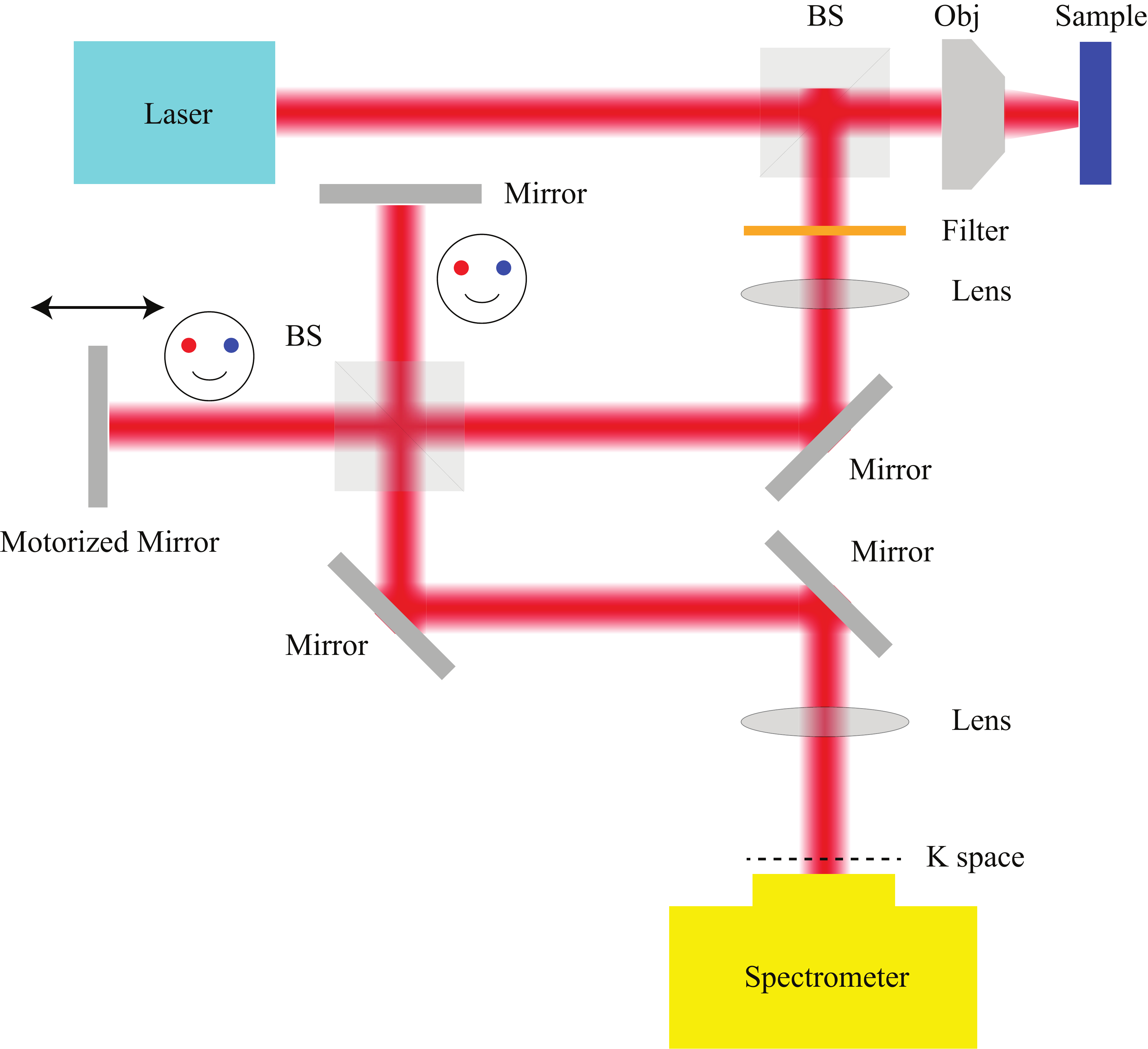}
    \caption{The time coherence of the emission from the sample was collected by an objective with the NA $\sim$ 0.5 and was sent to the Michelson interferometer. One of the light paths can be tuned by the one-dimensional translation stage and through a final lens to image it to the momentum space. Eventually, the signal was guided to the spectrometer coupled with a CCD. The “smiling face” represents the two interfering images.}
    \label{fig:S3}
\end{figure}

\subsection*{S4. Numerical Method}
To simulate the dynamics of the polaritons, we make use of the semiclassical Boltzmann equation, which reads for the lower and upper polariton branches: 
\begin{equation}
\begin{split} 
\frac{\partial n^{\mathrm{LP}}_{\vec{k}}}{\partial t}&=P^{\mathrm{LP}}_{\vec{k}}(t)-\frac{n^{\mathrm{LP}}_{\vec{k}}}{\tau^{\mathrm{LP}}_{\vec{k}}}+\sum_{\vec{k}^{\prime}} W_{\vec{k}^{\prime} \rightarrow \vec{k}}^{(i)}(t)-\sum_{\vec{k}^{\prime}} W_{\vec{k} \rightarrow \vec{k}^{\prime}}^{(i)}(t)\\
\frac{\partial n^{\mathrm{UP}}_{\vec{k}}}{\partial t}&=P^{\mathrm{UP}}_{\vec{k}}(t)-\frac{n^{\mathrm{UP}}_{\vec{k}}}{\tau^{\mathrm{UP}}_{\vec{k}}}+\sum_{\vec{k}^{\prime}} W_{\vec{k}^{\prime} \rightarrow \vec{k}}^{(i)}(t)-\sum_{\vec{k}^{\prime}} W_{\vec{k} \rightarrow \vec{k}^{\prime}}^{(i)}(t)
\end{split}
\end{equation}
Here  ${n_{\vec{k}}}$ is the occupation number, $\tau_{\vec{k}}$ is the characteristic lifetime, $P_{\vec{k}}$ is the pumping term and the label $\mathrm{LP}$ and $\mathrm{UP}$ refer to the lower and upper polariton branches respectively. The $W^{(i)}$ are the interaction terms for particle-particle collisions within each population. The values of $\vec{k}$ and $\vec{k'}$ range over the mesh in k-space. Each bin in the mesh holds the number of particles within the width of that bin.  The two branches (lower and upper polaritons) use the same mesh in k-space and each uses its respective dispersion curve given by:
\begin{equation}
\begin{split} 
E_{\mathrm{LP}, \mathrm{UP}}\left(k_{\|}\right)=\frac{1}{2}\left[E_{\mathrm{exc}}+E_{\mathrm{cav}} \pm \sqrt{\left(E_{\mathrm{exc}}-E_{\mathrm{cav}}\right)^2+\Omega^2}\right]
\end{split}
\end{equation}
Where $k_{\|}$ is  in-plane wave number, $\Omega$ is the Rabi splitting and $E_{\mathrm{exc}}$ and $E_{\mathrm{cav}}$ are the dispersions for the bare exciton and the cavity photon respectively, which are given by: 

\begin{equation}
\begin{split} 
E_{\mathrm{exc}} = E_{0}+\frac{\hbar^2k_{\|}^{2}}{2m_{\mathrm{exc}}}\\
E_{\mathrm{cav}}=\frac{\hbar c}{n_c} \sqrt{k_{\perp}^2+k_{\|}^2}
\end{split}
\end{equation}
Here $m_{\mathrm{exc}}$ mass of the exciton, $c$ is the speed of light and $k_{\perp}=n_c\left(2 \pi / \lambda_c\right)$ is the perpendicular wave number, where $\lambda_c$ is the resonance wavelength of the cavity and $n_c$ is the refractive index of the cavity. 
\par
\begin{figure}[!ht]
    \centering
    \includegraphics[width=0.7\textwidth]{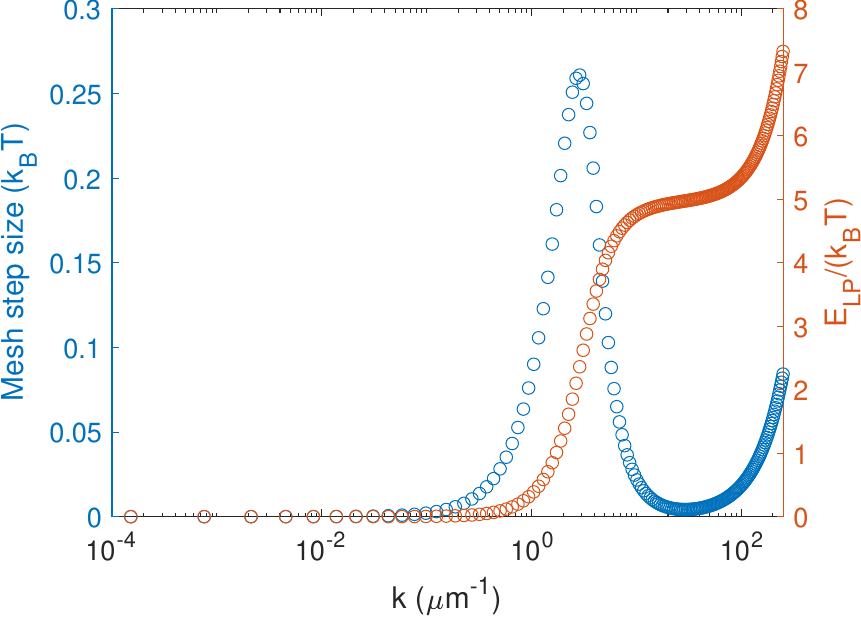}
    \caption{The energy step size as a function of wave number on the mesh. The dispersion curve for the lower polariton is plotted on the secondary vertical axis.}
    \label{fig:S4}
\end{figure}

For the mesh choice, we  follow the same approach as in Ref. \cite{hartwell2010numerical}. The mesh space is chosen such that there are many points near $k=0$, fewer points in the bottleneck region, and then many points again for the flat excitonic region. The largest bin in energy $\Delta E$ is chosen such that it is well below $k_{b}T$. Additionally, the highest energy in the mesh is chosen such that it is much larger than $k_{b}T$, typically $E_{\max } \sim 30 k_B T$, to ensure highest energy has a very low occupation number. A plot of the bin width chosen in the simulations is shown in Fig. \ref{fig:S4}. 
\par
The average occupations per state for the LP and UP branches is then calculated at each k-space bin according to the equation: 
\begin{equation}
\begin{split} 
\Delta n_k=\frac{\partial n_k}{\partial t} \Delta t
\end{split}
\end{equation}
We choose $\Delta t$ such that in each iteration $1\%$ of the total particles get moved around. The rate $\frac{\partial n_k}{\partial t}$ is described by Eq. (1). The updated $n_k(t)$ is then used to find the new $W_{\vec{k} \rightarrow \vec{k}^{\prime}}(t)$ until a steady-state distribution is reached. The laser generation is modeled by using a time-dependent pump term $P_{\vec{k}}(t)$. Since the laser generation is nonresonant with energy much greater than the polariton energies, we assumed that the free electrons and holes created in the pump process lead each polariton state being pumped with equal probability. The decay of the polaritons are updated in each time step as written in Eq. (1). The characteristic lifetime of the upper polariton and lower polariton branches is a function of the exciton and photon fraction and is given by:
\begin{equation}
\begin{split} 
\begin{aligned}
\frac{1}{\tau^{\mathrm{LP}}\left(k_{\|}\right)} & =\frac{\left|C^\mathrm{LP}_{k_{\|}}\right|^2}{\tau_{\mathrm{cav}}}+\frac{\left|X^\mathrm{LP}_{k_{\|}}\right|^2}{\tau_{\mathrm{nr}}} \\
\frac{1}{\tau^{\mathrm{UP}}\left(k_{\|}\right)} & =\frac{\left|C^\mathrm{UP}_{k_{\|}}\right|^2}{\tau_{\mathrm{cav}}}+\frac{\left|X^\mathrm{UP}_{k_{\|}}\right|^2}{\tau_{\mathrm{nr}}}
\end{aligned}
\end{split}
\end{equation}
Where $\tau_{\mathrm{cav}}$ is the cavity lifetime, $\tau_{\mathrm{nr}}$ is the non-radiative of the excitons. In the simulations, we used $\tau_{\mathrm{cav}} = 0.6\;\mathrm{ps}$ and $\tau_{\mathrm{nr}}=100\;\mathrm{ns}$. The Hopfield coefficients are given by \cite{hopfield1958theory},
\begin{equation}
\begin{split} 
\begin{aligned}
\left|X^\mathrm{LP}_{k_{\|}}\right|^2&=\frac{1}{2}\left(1+\frac{\Delta_{k_{\|}}}{\sqrt{\Delta_{k_{\|}}^2+\Omega^2}}\right)\\
\left|C^\mathrm{LP}_{k_{\|}}\right|^2&=\frac{1}{2}\left(1-\frac{\Delta_{k_{\|}}}{\sqrt{\Delta_{k_{\|}}^2+\Omega^2}}\right)\\
\left|X^\mathrm{UP}_{k_{\|}}\right|^2&=\frac{1}{2}\left(1-\frac{\Delta_{k_{\|}}}{\sqrt{\Delta_{k_{\|}}^2+\Omega^2}}\right)\\
\left|C^\mathrm{UP}_{k_{\|}}\right|^2&=\frac{1}{2}\left(1+\frac{\Delta_{k_{\|}}}{\sqrt{\Delta_{k_{\|}}^2+\Omega^2}}\right)
\end{aligned}
\end{split}
\end{equation}
where $\Delta_{k_{\|}}=E_{\mathrm{cav}}\left(k_{\|}\right)-E_{\text {exc }}\left(k_{\|}\right)$ is the detuning. 
\par
As mentioned in the main text, our model includes four types of interactions: polariton-polariton interactions within the same branch, polariton-polariton interactions between different branches (i.e upper polariton-lower polariton interaction), polariton-phonon interactions, and polariton-electron interactions. In the following sections, we list all the scattering rates used in our model.
\subsubsection*{Lower polartion-lower polariton interactions}
The net out-scattering rate of a state $\vec{k}$ an be calculated using Fermi's golden rule:
\begin{equation}
\begin{split} 
\begin{aligned}
\frac{d n_{\vec{k}}}{d t}=\frac{2 \pi}{\hbar} \sum_{\vec{k}_{f}}\left|\left\langle\vec{k}_{f}\left|V_{\text {int }}\right| \vec{k}_{i}\right\rangle\right|^{2} \delta\left(E_{\text {f}}-E_{\text {i}}\right)
\end{aligned}
\end{split}
\end{equation}
where two lower polariton particles scatter from the state $|\vec{k}_{i}\rangle=\left|\vec{k} \vec{k}_3\right\rangle$ to $|\vec{k}_{f}\rangle=\left|\vec{k}_{1} \vec{k}_2\right\rangle$. The interaction term $V_\text {int}$ for this two-body scattering process can be written as: 
\begin{equation}
\begin{split} 
\begin{aligned}
V_{i n t}=|M| a_{\vec{k}_1}^{\dagger} a_{\vec{k}_2}^{\dagger} a_{\vec{k}_3} a_{\vec{k}},
\end{aligned}
\end{split}
\end{equation}
where $a_{\vec{k}}^{\dagger}$ and $a_{\vec{k}}$ are the bosonic creation and destruction operators for polaritons and $M$ is the matrix element. Putting the two-body interaction term into Fermi's golden rule gives:
\begin{equation}
\begin{split} 
\begin{aligned}
\frac{\partial n_{\vec{k}}}{\partial t}= & \frac{2 \pi}{\hbar} \sum_{\vec{k}_1 \vec{k}_2}\left|M\left(\left|\vec{k}-\vec{k}_2\right|\right)\right|^2 n_{\vec{k}} n_{\vec{k}_3} \left (1+n_{\vec{k}_1}  \right )\left (1+n_{\vec{k}_2}  \right ) \\
& \times \delta \left (E_\mathrm{LP}(\vec{k})+E_\mathrm{LP}(\vec{k}_3)-E_\mathrm{LP}(\vec{k}_2)-E_\mathrm{LP}(\vec{k}_1)  \right ),
\end{aligned}
\end{split}
\end{equation}
In the thermodynamic limit, the sums can be converted to an integral, which gives: 
\begin{equation}
\begin{split} 
\begin{aligned}
\frac{\partial n_{\vec{k}}}{\partial t}= & \frac{S^2}{(2 \pi)^3 \hbar} \int d^2 \vec{k}_1 d^2 \vec{k}_2\left|M\left(\left|\vec{k}-\vec{k}_2\right|\right)\right|^2 n_{\vec{k}} n_{\vec{k}_3}\left (1+n_{\vec{k}_1}   \right )\\
& \times\left (1+n_{\vec{k}_2}   \right )\delta\left (E_\mathrm{LP}(\vec{k})+E_\mathrm{LP}(\vec{k}_3)-E_\mathrm{LP}(\vec{k}_2)-E_\mathrm{LP}(\vec{k}_1)  \right ),
\end{aligned}
\end{split}
\end{equation}
Where $S$ is the area of the sample. $\vec{k}_3$ can be eliminated by momentum conservation, which insures $\vec{k}_3+\vec{k}=\vec{k}_1+\vec{k}_2$. Additionally, We assume that the scattering rate is independent of angle in k-space $n (\vec{k}) = n ( | \vec{k}  | )$. A similar experstion can be found for the in-scattering rate by replacing the statistical factors,
\begin{equation}
\begin{split} 
\begin{aligned}
n_{\vec{k}} n_{\vec{k}_3}\left (1+n_{\vec{k}_1}   \right )\left (1+n_{\vec{k}_2}   \right )\rightarrow n_{\vec{k_1}} n_{\vec{k}_2}\left (1+n_{\vec{k}}   \right )\left (1+n_{\vec{k}_3}   \right )
\end{aligned}
\end{split}
\end{equation}
An approximation for the matrix element $M$ was calculated by Tassone and
Yamamoto \cite{tassone1999exciton} to be,
\begin{equation}
\begin{split} 
\begin{aligned}
M \sim 6 X^\mathrm{LP}_{\vec{k}} X^\mathrm{LP}_{\vec{k}_1} X^\mathrm{LP}_{\vec{k}_2} X^\mathrm{LP}_{\vec{k}_3} E_B \frac{a_B^2}{S} .
\end{aligned}
\end{split}
\end{equation}
where $a_B \approx 1\; \mathrm{nm}$ \cite{chakraborty2018control} is the exciton Bhor radius and $E_B \approx 0.7\; \mathrm{eV}$ \cite{mak2016photonics} is the binding energy of the exciton for WS$_2$.
\subsubsection*{Upper polartion-upper polariton interactions}
The upper polariton- upper polariton interaction takes the same form as the lower polariton-lower polariton interactions  but with the substitution $E_\mathrm{LP}\rightarrow E_\mathrm{UP}$ and $X^\mathrm{LP}\rightarrow X^\mathrm{UP}$. Therefore, the out-scattering rate is given by: 
\begin{equation}
\begin{split} 
\begin{aligned}
\frac{\partial n_{\vec{k}}}{\partial t}= & \frac{S^2}{(2 \pi)^3 \hbar} \int d^2 \vec{k}_1 d^2 \vec{k}_2\left|M\left(\left|\vec{k}-\vec{k}_2\right|\right)\right|^2 n_{\vec{k}} n_{\vec{k}_3}\left (1+n_{\vec{k}_1}   \right )\\
& \times\left (1+n_{\vec{k}_2}   \right )\delta\left (E_\mathrm{UP}(\vec{k})+E_\mathrm{UP}(\vec{k}_3)-E_\mathrm{UP}(\vec{k}_2)-E_\mathrm{UP}(\vec{k}_1)  \right ),
\end{aligned}
\end{split}
\end{equation}
and
\begin{equation}
\begin{split} 
\begin{aligned}
M \sim 6 X^\mathrm{UP}_{\vec{k}} X^\mathrm{UP}_{\vec{k}_1} X^\mathrm{UP}_{\vec{k}_2} X^\mathrm{UP}_{\vec{k}_3} E_B \frac{a_B^2}{S} .
\end{aligned}
\end{split}
\end{equation}
\subsubsection*{Lower polartion-upper polariton interactions}
In additon to lower polariton-lower polariton and upper polariton-upper polariton interactions, our model also takes into account the lower polariton-upper polariton interactions. In this case, a lower polariton particles scatters with an upper polariton particle from the state $|\vec{k}_{i}\rangle=\left|\vec{k} \vec{k}_3\right\rangle$ to $|\vec{k}_{f}\rangle=\left|\vec{k}_{1} \vec{k}_2\right\rangle$. Where $\vec{k}$ and $\vec{k}_1$ is the incoming and ourcoming lower polariton wavevectors respectively, and $\vec{k}_3$ and $\vec{k}_2$ is the incoming and ourcoming upper polariton wavevectors respectively. Following the same procedure as in the previous sections, the lower polariton out-scattering rate is written as: 
\begin{equation}
\begin{split} 
\begin{aligned}
\frac{\partial n^\mathrm{LP}_{\vec{k}}}{\partial t}= & \frac{S^2}{(2 \pi)^3 \hbar} \int d^2 \vec{k}_1 d^2 \vec{k}_2\left|M\left(\left|\vec{k}-\vec{k}_2\right|\right)\right|^2 n_{\vec{k}} n_{\vec{k}_3}\left (1+n_{\vec{k}_1}   \right )\\
& \times\left (1+n_{\vec{k}_2}   \right )\delta\left (E_\mathrm{LP}(\vec{k})+E_\mathrm{UP}(\vec{k}_3)-E_\mathrm{UP}(\vec{k}_2)-E_\mathrm{LP}(\vec{k}_1)  \right ),
\end{aligned}
\end{split}
\end{equation}
and
\begin{equation}
\begin{split} 
\begin{aligned}
M \sim 6 X^\mathrm{LP}_{\vec{k}} X^\mathrm{LP}_{\vec{k}_1} X^\mathrm{UP}_{\vec{k}_2} X^\mathrm{UP}_{\vec{k}_3} E_B \frac{a_B^2}{S} .
\end{aligned}
\end{split}
\end{equation}
\par
These calculations assume that there are no lower polariton to upper polariton conversion or vice versa during the two-body scattering. This is a reasonable assumption since Rabi splitting is much larger than the energy exchange in the two-body collision for any resonable wavevector $k$.
\par
The same results is obtained for the upper polariton out-scattering rate,
\begin{equation}
\begin{split} 
\begin{aligned}
\frac{\partial n^\mathrm{UP}_{\vec{k}_3}}{\partial t}= & \frac{S^2}{(2 \pi)^3 \hbar} \int d^2 \vec{k}_1 d^2 \vec{k}_2\left|M\left(\left|\vec{k}-\vec{k}_2\right|\right)\right|^2 n_{\vec{k}} n_{\vec{k}_3}\left (1+n_{\vec{k}_1}   \right )\\
& \times\left (1+n_{\vec{k}_2}   \right )\delta\left (E_\mathrm{LP}(\vec{k})+E_\mathrm{UP}(\vec{k}_3)-E_\mathrm{UP}(\vec{k}_2)-E_\mathrm{LP}(\vec{k}_1)  \right ),
\end{aligned}
\end{split}
\end{equation}

\subsubsection*{Lower polartion-longitudinal acoustic phonon interactions}
The polaritons in consideration are two-dimensional particles with in-plane dispersion and they scatter with acoustic phonons that are three-dimensional. This scattering process conserves the in-plane momentum. The phonon wavevector can be written as  $\vec{q}=\left(\vec{q}_{\|}, \vec{q}_{z}\right)$, where $\vec{q}_{\|}$ and $\vec{q}_{z}$ are the in-plane and the $z$-component (growth direction) of the phonon wavevector respectively. When considering the polariton-phonon interaction, there are 4 different processes. These processes are 1) scattering out of state $\vec{k}$ while emitting a phonon 2) scattering out of state $\vec{k}$ while absorbing a phonon 3) scattering into a state $\vec{k}$ while emitting a phonon 4) scattering into state $\vec{k}$ while absorbing a phonon. The polariton-phonon interaction is based on hydrostatic deformation potential, which takes the form \cite{piermarocchi1996nonequilibrium}:
\begin{equation}
\begin{split} 
\begin{aligned}
\begin{aligned}
M(\vec{k}, \vec{q})= & i X^\mathrm{LP}_{k} X^\mathrm{LP}_{k^{\prime}}\sqrt{\frac{\hbar\left(q_{\|}^2+q_z^2\right)^{1 / 2}}{2 \rho V u}} \left[a_e I_e^{\|}\left((|\vec{q}|) I_e^{\perp}\left(q_z\right)-a_h I_h^{\|}(|\vec{q}|) I_h^{\perp}\left(q_z\right)\right]\right.
\end{aligned}
\end{aligned}
\end{split}
\end{equation}
Where $V$, $\rho$, $u$ are volume, density and longitudinal sound velocity respectively. $a_{e} = -11.62\;\mathrm{eV}$ and $a_{h} = -5.18\;\mathrm{eV}$ \cite{rawat2018comprehensive} are the deformation coefficients of the conduction and valence band for WS$_2$ respectively. The density and averaged longitudinal sound velocity for WS$_2$ are taken to be $\rho = 7.5 \;\mathrm{g/cm^{3}}$ and $u = 5953\;\mathrm{m/s}$ \cite{persson2011phononic}. $I_{\mathrm{e}(\mathrm{h})}^{\perp(\|)}$ are the overlap integrals between the exciton and phonon mode. These overlap integrals are given by \cite{tassone1997bottleneck,andreani1990accurate,bastard1986electronic,pau1995microcavity,piermarocchi1996nonequilibrium}: 
\begin{equation}
\begin{split} 
\begin{aligned}
\begin{aligned}
\begin{gathered}
I_{e(h)}^{\perp}\left(q_z\right)=\frac{8 \pi^2}{L_z q_z\left(4 \pi^2-L_z^2 q_z^2\right)} \sin \left(\frac{L_z q_z}{2}\right) \\
I_{e(h)}^{\|}=\left[1+\left(\frac{m_{h(e)}}{2 M}\left|q_{\|}\right| a_B\right)^2\right]^{-3 / 2},
\end{gathered}
\end{aligned}
\end{aligned}
\end{split}
\end{equation}
where $L_{z} \approx 0.7 \;\mathrm{nm}$ \cite{mak2016photonics} is the WS$_2$ monolayer thickness and $a_{B}$ is the two-dimensional exciton Bohr radius. $m_{h(e)}$ is the electorn (hole) mass and is taken to be $m_e = m_h = 0.4m^{\mathrm{vacc}}_e$ \cite{mak2016photonics}, where $m^{\mathrm{vacc}}_e$ is the vaccume electron mass. 
\par
The total scattering rate is given by:
\begin{equation}
\begin{split} 
\begin{aligned}
\begin{gathered}
\begin{aligned}
\frac{d n_{\vec{k}}}{d t}= & -\frac{2 \pi}{\hbar} \sum_{\vec{k}_1, \vec{q}_{z}}|M(\vec{k}, \vec{q})|^2\left[n_{\vec{q}}^{\mathrm{phon}}+\frac{1}{2} \pm \frac{1}{2}\right] n_{\vec{k}}\left(1+n_{\vec{k}_1}\right) \delta\left(E_\mathrm{LP}(\vec{k}_1)-E_\mathrm{LP}(\vec{k}) \pm \hbar \nu \vec{q}\right) \\
& +\frac{2 \pi}{\hbar} \sum_{\vec{k}_1, \vec{q}_{z}}|M(\vec{k}, \vec{q})|^2\left[n_{\vec{q}}^{\text {phon }}+\frac{1}{2} \pm \frac{1}{2}\right] n_{\vec{k}_1}\left(1+n_{\vec{k}}\right) \delta\left(E_\mathrm{LP}(\vec{k})-E_\mathrm{LP}(\vec{k}_1) \pm \hbar \nu \vec{q}\right)
\end{aligned}
\end{gathered}
\end{aligned}
\end{split}
\end{equation}
Where $\pm$ correpsond to phonon emission $(+)$ and phonon absorption $(-)$. The conservation of in-plane momentum insures that $\vec{q}_{\|}=\vec{k}-\vec{k}_1$. Here, we assume that the phonons are in thermal equilibrium with $n_{\vec{q}}^{\text {phon }}$ given by the Planck distribution $n_{\vec{q}}^{\text {phon }} = 1/\left (e^{\hbar \omega_{q}/k_{B}T}-1  \right )$. 
\subsubsection*{Upper polartion-longitudinal acoustic phonon interactions}
The upper polariton-longitudinal acoustic phonon interaction takes the same form as the lower polariton-longitudinal acoustic phonon interaction but with the substitution $E_\mathrm{LP}\rightarrow E_\mathrm{UP}$ and $X^\mathrm{LP}\rightarrow X^\mathrm{UP}$. Therefore, the total scattering rate is given by:
\begin{equation}
\begin{split} 
\begin{aligned}
\begin{gathered}
\begin{aligned}
\frac{d n_{\vec{k}}}{d t}= & -\frac{2 \pi}{\hbar} \sum_{\vec{k}_1, \vec{q}_{z}}|M(\vec{k}, \vec{q})|^2\left[n_{\vec{q}}^{\mathrm{phon}}+\frac{1}{2} \pm \frac{1}{2}\right] n_{\vec{k}}\left(1+n_{\vec{k}_1}\right) \delta\left(E_\mathrm{UP}(\vec{k}_1)-E_\mathrm{UP}(\vec{k}) \pm \hbar \nu \vec{q}\right) \\
& +\frac{2 \pi}{\hbar} \sum_{\vec{k}_1, \vec{q}_{z}}|M(\vec{k}, \vec{q})|^2\left[n_{\vec{q}}^{\text {phon }}+\frac{1}{2} \pm \frac{1}{2}\right] n_{\vec{k}_1}\left(1+n_{\vec{k}}\right) \delta\left(E_\mathrm{UP}(\vec{k})-E_\mathrm{UP}(\vec{k}_1) \pm \hbar \nu \vec{q}\right)
\end{aligned}
\end{gathered}
\end{aligned}
\end{split}
\end{equation}
with,
\begin{equation}
\begin{split} 
\begin{aligned}
\begin{aligned}
M(\vec{k}, \vec{q})= & i X^\mathrm{UP}_{k} X^\mathrm{UP}_{k^{\prime}}\sqrt{\frac{\hbar\left(q_{\|}^2+q_z^2\right)^{1 / 2}}{2 \rho V u}} \left[a_e I_e^{\|}\left((|\vec{q}|) I_e^{\perp}\left(q_z\right)-a_h I_h^{\|}(|\vec{q}|) I_h^{\perp}\left(q_z\right)\right]\right.
\end{aligned}
\end{aligned}
\end{split}
\end{equation}
\subsubsection*{Upper polariton-lower polariton conversion via optical phonons}
In our model, the only process that converts an upper polariton to a lower polariton and vice versa is the optical phonon interaction. The polariton-optical phonon interaction is based on the Frohlich interaction, which takes the form
\begin{equation}
\begin{split} 
\begin{aligned}
\begin{aligned}
\begin{aligned}
M(|\vec{k}, \vec{q}|)= & i X_{k} X_{k^{\prime}} \sqrt{\frac{2 \pi e^2 \hbar \omega_{L O}}{\left(\vec{q}_{\|}^2+q_z^2\right) V}}\left(\frac{1}{\epsilon_{\infty}}-\frac{1}{\epsilon_0}\right)^{1 / 2} \\
& \times\left[a_e I_e^{\|}(|\vec{q}|) I_e^{\perp}\left(q_z\right)-a_h I_h^{\|}(|\vec{q}|) I_h^{\perp}\left(q_z\right)\right]
\end{aligned}
\end{aligned}
\end{aligned}
\end{split}
\end{equation}
with,
\begin{equation}
\begin{split} 
\begin{aligned}
\begin{aligned}
\begin{gathered}
I_{e(h)}^{\perp}\left(q_z\right)=\frac{8 \pi^2}{L_z q_z\left(4 \pi^2-L_z^2 q_z^2\right)} \sin \left(\frac{L_z q_z}{2}\right) \\
I_{e(h)}^{\|}=\left[1+\left(\frac{m_{h(e)}}{2 M}\left|q_{\|}\right| a_B\right)^2\right]^{-3 / 2},
\end{gathered}
\end{aligned}
\end{aligned}
\end{split}
\end{equation}
For a scattering out of state $\vec{k}_{UP}$ to a state $\vec{k}_{LP}$ while emitting an optical phonon, we have:
\begin{equation}
\frac{\partial n_{\vec{k}_\mathrm{UP}}}{\partial t}=\frac{2 \pi}{\hbar} \sum_{\vec{k}_\mathrm{LP}, \vec{q}_z}|M|^2 n_{\vec{k}_\mathrm{UP}}\left[1+n_{\vec{k}_{LP}}\right]\left[1+n_{\vec{q}}\right] \delta\left(E_\mathrm{LP}(\vec{k}_\mathrm{LP})+E(\vec{q})-E_\mathrm{UP}(\vec{k}_\mathrm{UP})\right)
\end{equation}
Since the optical phonon has a flat dispersion, we can approximate its energy as constant $E(\vec{q})= \hbar \omega_{LO}$, with $\omega_{LO}  \approx 1.05\times 10^{13} \;\mathrm{Hz}$ for WS$_2$ \cite{molina2011phonons}.
\subsubsection*{Electron-polariton interactions}
The electron-polariton interaction takes into account both direct and exchange interactions. The interaction Hamiltonian for the electron-electron exchange is given by:
\begin{equation}
H_{e x}=\frac{1}{2 S} \frac{e^2}{\epsilon(|\Delta k|+\kappa)} c_{k-\Delta k}^{\dagger} c_{k+\Delta k}^{\dagger} c_k c_k
\end{equation}
where $c_{k-\Delta k}^{\dagger}$ and $c_{k-\Delta k}$ are the electron creation and destruction operators respectively, $\Delta k$ is the momentum exchange, $e$ is the electron charge and $\kappa$ is the screening parameter. In two-dimensions, the Debye-Huckel screening parameter takes the form:
\begin{equation}
\kappa=\frac{\mathrm{e}^2}{2 \epsilon_{\infty}} \frac{n}{k_B T}
\end{equation}
where $n$ is the density, $T$ is the electron temperature and $k_B$ is the Boltzmann constant. The matrix element for the polariton electron interaction is then given by $M=\left\langle f\left|H_{e x}\right| i\right\rangle$. This matrix element is then placed into Fermi's Golden rule as was done in the previous sections to calculate the lower polartion-electron and upper polariton-electron interactions. For more details on the derivation of the matrix element, we refer the reader to Ref. \cite{hartwell2010numerical}. 
\printbibliography
\end{document}